\documentclass[%
 reprint,
superscriptaddress,
nofootinbib,
 amsmath,amssymb,
 aps,
pra,
]{revtex4-2}

\usepackage[utf8]{inputenc}
\usepackage[english]{babel}

\usepackage{amsmath}
\usepackage{amssymb}
\usepackage{bbm}
\usepackage{bm}
\usepackage{braket}
\usepackage{graphicx}
\usepackage{dsfont}

\usepackage[usenames,dvipsnames]{xcolor}
\usepackage[colorlinks=true,citecolor=MidnightBlue,linkcolor=MidnightBlue,urlcolor=MidnightBlue]{hyperref}

\usepackage[capitalize]{cleveref}
 
\begin{document}

\title{Heralding Higher-Dimensional Bell and Greenberger-Horne-Zeilinger States Using Multiport Splitters}

\author{Daniel Bhatti}
\affiliation{Networked Quantum Devices Unit, Okinawa Institute of Science and Technology Graduate University, Okinawa, Japan}
\affiliation{Institute for Functional Matter and Quantum Technologies, University of Stuttgart, 70569 Stuttgart, Germany}
\affiliation{Center for Integrated Quantum Science and Technology (IQST), University of Stuttgart, 70569 Stuttgart, Germany}
\author{Stefanie Barz}
\affiliation{Institute for Functional Matter and Quantum Technologies, University of Stuttgart, 70569 Stuttgart, Germany}
\affiliation{Center for Integrated Quantum Science and Technology (IQST), University of Stuttgart, 70569 Stuttgart, Germany}

\begin{abstract}
One of the most important resources for quantum optical experiments and applications are on-demand highly entangled multiphoton quantum states. A promising way of generating them is heralding entanglement generation at a high rate from letting independent photons interfere. However, such schemes often work for a specific internal degree of freedom of the interfering photons only. Going to higher numbers of entangled photons, the success probabilities decrease while the number of necessary resources, e.g., auxiliary photons and optical elements, increases. To make probabilistic schemes feasible also for larger quantum states, it is therefore important to find resource-efficient generation schemes with high success probabilities. In this work, we introduce easily implementable schemes to herald qubit Greenberger-Horne-Zeilinger (GHZ) states, higher-dimensional Bell states and higher-dimensional three-party GHZ states. Our schemes solely rely on multiphoton interference, i.e., they can be adjusted to work for arbitrary degrees of freedom. Furthermore, they demonstrate high success probabilities and need comparably few auxiliary photons.
\end{abstract}

\maketitle

\section{Introduction}

Highly entangled quantum states have various important applications ranging from the field of quantum communication~\cite{Ekert1991,Hillery1999,Bai2017,Murta2020,Erhard2020}, to testing the fundamentals of quantum mechanics~\cite{Vertesi2010,Tang2013,Erhard2018Experiment}. 
Furthermore, when having access to Bell and Greenberger-Horne-Zeilinger (GHZ) states, one can generate arbitrary graph states by utilizing fusion protocols~\cite{Gimeno-Segovia2015,Paesani2021}. This enables fusion-based quantum computing~\cite{Wang2020,Bartolucci2023}.

The generation of entangled photonic qubit states has intensively been studied~\cite{Sagi2003,Lim2005,Pilnyak2017,Bhatti2023,Wang2018,Kumar2023,Meyer-Scott2022,Lindner2009,Economou2010,Gimeno-Segovia2019,Besse2020,Thomas2022,Gubarev2020,Zhang2008,Gimeno-Segovia2015Thesis,Paesani2021,Bartolucci2021,Shimizu2024,Maring2023,Cao2024,Chen2024}. Probabilistic state specific schemes based on postselection in principle allow for the generation of arbitrarily large states~\cite{Sagi2003,Lim2005,Pilnyak2017,Bhatti2023}. However, the success probabilities exponentially decrease with increasing numbers of photons, limiting the generated states to small numbers of qubits or low rates~\cite{Wang2018,Kumar2023}. By utilizing multiplexing and feed forward, it has been shown that high generation rates can be reached for entangled states with up to six qubits~\cite{Meyer-Scott2022}.
Although postselecting entangled states in many current scenarios is sufficient it is necessary to have on-demand generated quantum states, e.g., in quantum teleportation, but especially for fusing arbitrary graph states~\cite{Gimeno-Segovia2015}.

One approach to generate entangled states on-demand is to use deterministic single-photon emitters. In addition to the theoretical investigations~\cite{Lindner2009,Economou2010,Gimeno-Segovia2019}, deterministic qubit graph states have been generated experimentally~\cite{Besse2020,Thomas2022}. Furthermore, even the fusion of deterministically generated photonic graph states has been demonstrated~\cite{Thomas2024}. However, these schemes require sophisticated experimental setups and are therefore very expensive~\cite{Cao2024}.

A different approach is to probabilistically herald entanglement generation using only linear optical elements~\cite{Zhang2008,Gimeno-Segovia2015Thesis,Paesani2021,Bartolucci2021,Shimizu2024,Maring2023,Cao2024,Chen2024}.
Such schemes typically require many auxiliary photons and optical elements. At the same time, they suffer from the same decrease in success probability as the post-selection schemes, what makes them suitable especially for small numbers of photons~\cite{Zhang2008,Maring2023,Cao2024,Chen2024}. Additionally, they often work for a specific degree of freedom only, e.g., polarization encoding~\cite{Zhang2008,Gimeno-Segovia2015Thesis}, or path encoding~\cite{Paesani2021,Bartolucci2021,Shimizu2024}.
Recently, different experimental results have demonstrated heralded entanglement generation using linear optics~\cite{Maring2023,Cao2024,Chen2024}. All these experiments focus on the generation of three-qubit GHZ states, which is sufficient for producing arbitrary graph states~\cite{Gimeno-Segovia2015}.

Instead of increasing the number of qubits, it can be beneficial to use entangled higher-dimensional qudit states. Employing qudit states allows for a denser encoding of information, increased noise tolerance and simplification of experimental setups~\cite{Wang2020,Erhard2020,Friis2019,Goel2024}. Postselecting qudit Bell states can be achieved from using multiple probabilistic photon sources, what has been shown theoretically and experimentally~\cite{Schaeff2012,Jianwei2018}. For postselecting qudit GHZ states, only a few results have been published, limited in the number of photons~\cite{Erhard2018Experiment,RuizGonzalez2023}. 
Recently, it has been shown that arbitrary entangled qudit states can be generated from using linear optics only~\cite{Paesani2021}. However, such schemes possess low success probabilities, and need either many additional auxiliary modes and photons or many optical elements~\cite{Paesani2021,Chin2024}.
Besides the heralding schemes, also theoretical work has been published demonstrating how to use deterministic quantum emitters to generate qudit graph and GHZ states, which again require sophisticated experimental setups~\cite{Bell2022,Raissi2024}.

In this work we focus on using symmetric multiport beam splitters (SMSs), i.e., linear optical elements, and present three different easily implementable schemes to herald entanglement generation. i) In \cref{sec:Bell4portSMS}, we start by presenting a scheme based on four-port SMSs (4SMSs) to herald the generation of qubit GHZ states of arbitrary size, which works for arbitrary internal degrees of freedom. ii) In \cref{sec:Bell3SMS}, we present a scheme based on three-port SMSs (3SMS) to herald the generation of qudit Bell states for arbitrary dimensions. The scheme demonstrates high success probabilities, and needs comparably few numbers of auxiliary modes, photons, and optical elements. It can be adjusted to work for arbitrary internal degrees of freedom. iii) In \cref{sec:GHZ4SMS}, we extend the previous scheme by exchanging all 3SMSs by 4SMSs, now heralding three-party qudit GHZ states. It has the same advantages as the previous scheme. In \cref{sec:PhotonSubtraction}, we then show how deterministic photon subtraction as used in Ref.~\cite{Bartolucci2021} can increase the success probabilities of schemes ii) and iii) even further.

\section{Background}
\label{sec:Theory}

\begin{figure}[t]
	\centering
		\includegraphics[width=0.85\columnwidth]{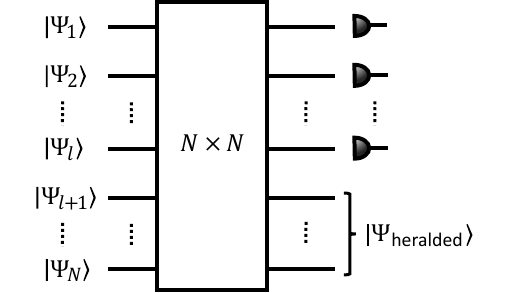}
	\caption{Unitary $N\times N$ network considered heralding Bell and GHZ states. Photonic input states $\ket{\Psi_i}$ ($i=1,2,\ldots,N$) are sent into the $N$ input modes of the $N$-port network. Detecting for specific photon-number distributions in particular output modes leads to heralded states $\ket{\Psi_{\text{heralded}}}$ in the remaining output modes.}
	\label{fig:SetupNetwork}
\end{figure}

In this work, we discuss different multiphoton states impinging on unitary $N$-port SMSs (see Fig.~\ref{fig:SetupNetwork}).
We show that by combining multiple SMSs and detecting specific photon-number distributions at certain output modes, the generation of higher-dimensional entangled states in the remaining modes can be heralded. We start by analyzing the two types of multiphoton input states important for this work.

The first input state of interest consists of one photon per input mode \cite{Lim2005IOP}
\begin{align} \label{eq:PsiIn}
	\ket{\Psi_{N,\text{in}}}  = \prod_{k=1}^{N} a_{F,k}^{\dagger}  \ket{0} & = \ket{1}_{F,1}\otimes \ket{1}_{F,2} \otimes \cdots \otimes \ket{1}_{F,N} \nonumber \\
 & \equiv \ket{1,1,\ldots,1}_{F},
\end{align}
where $a_{F,k}^{\dagger}$ ($a_{F,k}$) denotes the creation (annihilation) operator of a photon in the $k$th spatial input mode with the internal degree of freedom $F$, and $\ket{n}_{F,k}$ denotes the corresponding photon-number state in the Fock basis. Since here, we discuss the generation of $d$-dimensional quantum states we need photonic degrees of freedom capable of implementing a qudit system, i.e., $F \in \{0,1,\ldots,d \}$. For example, one could encode in path, frequency, time-bin, or orbital angular momentum degrees of freedom~\cite{Paesani2021,Borghi2023,Kysela2020,Zheng2022}.

The transformation of the $N$-mode input state to an $N$-mode output state using an $N$-port SMS can be described by a unitary matrix $U$ with matrix elements $U_{kl}$ ($k,l=1,2,\ldots,N$). The transition relation mapping input creation operators $a_{F,k}^{\dagger}$ to output creation operators $b_{F,l}^{\dagger}$ is given by~\cite{Bhatti2023}
\begin{align} \label{eq:TransformationRelation}
	a_{F,k}^{\dagger} \rightarrow \sum_{l=1}^{N} U_{kl} b_{F,l}^{\dagger}  .
\end{align} 

In Eq.~(\ref{eq:TransformationRelation}), the matrix elements $U_{kl}$ denote the transition for an input photon going from input mode $k$ to output mode $l$. For SMSs, those are defined by a discrete Fourier transform~\cite{Lim2005IOP}
\begin{align}
	U_{kl} = \frac{1}{\sqrt{N}} \omega_{N}^{(k-1)(l-1)} ,
\label{eq:Fourier}
\end{align}
with $\omega_{N}=\exp\!\left(\text{i} 2\pi/N\right)$ being the $N$th root of unity. The important multiports for this work are the 3SMS and the 4SMS.
Their matrices take the forms~\cite{Zukowski1997}
\begin{align}
\label{eq:tritter}
   U_{3} = \frac{1}{\sqrt{3}} \begin{pmatrix}
1 & 1 & 1\\
1 & e^{ \text{i} \frac{2 \pi}{3}  } & e^{\text{i} \frac{4 \pi}{3} } \\
1 & e^{\text{i} \frac{4 \pi}{3}} & e^{\text{i} \frac{8 \pi}{3}} 
\end{pmatrix} ,
\end{align}
and~\cite{Lim2005}
\begin{align}
    U_{4} = \frac{1}{2} \begin{pmatrix}
1 & 1 & 1 & 1\\
1 & \text{i} & -1 & -\text{i} \\
1 & -1 & 1 & -1 \\
1 & -\text{i} & -1 & \text{i} 
\end{pmatrix} ,
\end{align}
which allow us to calculate the complete output states for $N=3$:
\begin{align}
\label{eq:PsiOutTritter}
    \ket{\Psi_{3, \text{out}}} & =  \frac{\sqrt{2}}{3} \left( \ket{3,0,0}_{F} + \ket{0,3,0}_{F} + \ket{0,0,3}_{F} \right) \nonumber \\ 
    & \phantom{={}} - \frac{1}{\sqrt{3}} \ket{1,1,1}_{F}.
\end{align}
and for $N=4$:
\begin{align}
\label{eq:PsiOutQuitter}
    \ket{\Psi_{4, \text{out}}} & =  \frac{\sqrt{3}}{4\sqrt{2}} \left( \ket{4,0,0,0}_{F} - \ket{0,4,0,0}_{F} \right. \nonumber \\
    & \phantom{={}} \left. + \ket{0,0,4,0}_{F} - \ket{0,0,0,4}_{F} \right) \nonumber \\
    & \phantom{={}} + \frac{1}{2\sqrt{2}} \left( \ket{1,2,1,0}_{F} - \ket{0,1,2,1}_{F} \right. \nonumber \\
    & \phantom{={}} \left. + \ket{1,0,1,2}_{F} - \ket{2,1,0,1}_{F} \right) \nonumber \\
    & \phantom{={}} + \frac{1}{4} \left( \ket{0,2,0,2}_{F} - \ket{2,0,2,0}_{F} \right).
\end{align}

The second type of important input states consists of one photon per every second input mode. This works only for even $N$, i.e., in our case for $N=4$. The two possible input states are:
\begin{align}
    \ket{\Psi_{1-3, \text{in}}} = \ket{1,0,1,0}_{F},
\end{align}
and
\begin{align}
    \ket{\Psi_{2-4, \text{in}}} = \ket{0,1,0,1}_{F},
\end{align}
with the respective output states:
\begin{align}
\label{eq:Psi1-3}
    \ket{\Psi_{1-3, \text{out}}} & =  \frac{1}{2\sqrt{2}} \left( \ket{2,0,0,0}_{F} - \ket{0,2,0,0}_{F} \right. \nonumber \\
    & \phantom{={}} \left. + \ket{0,0,2,0}_{F} - \ket{0,0,0,2}_{F} \right) \nonumber \\
    & \phantom{={}} + \frac{1}{2} \left( \ket{1,0,1,0}_{F} - \ket{0,1,0,1}_{F} \right) ,
\end{align}
and
\begin{align}
\label{eq:Psi2-4}
    \ket{\Psi_{2-4, \text{out}}} & =  \frac{1}{2\sqrt{2}} \left( \ket{2,0,0,0}_{F} + \ket{0,2,0,0}_{F} \right. \nonumber \\
    & \phantom{={}} \left. + \ket{0,0,2,0}_{F} + \ket{0,0,0,2}_{F} \right) \nonumber \\
    & \phantom{={}} - \frac{1}{2} \left( \ket{1,0,1,0}_{F} + \ket{0,1,0,1}_{F} \right) .
\end{align}

Using these results, we can now discuss our heralding schemes.

\section{Heralding Qubit Bell and GHZ States Using 4-port Splitters}
\label{sec:Bell4portSMS}

The first setup is based on a single 4SMS and heralds a qubit Bell state. As we will show, this scheme can easily be extended to herald GHZ states of arbitrary size. The heralding mechanism solely relies on multiphoton interference and works independently of the internal degree of freedom chosen to encode the photonic qubit states $\ket{\mu}$ and $\ket{\eta}$. For example, one could use polarization or time-bin degrees of freedom.

\begin{figure}
    \centering
    \includegraphics[width=0.85\linewidth]{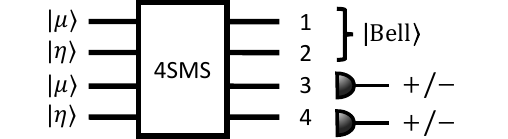}
    \caption{Setup to herald two-photon Bell states ($\ket{\text{Bell}}$) using a 4-port symmetric multiport splitter (4SMS). Four independent photons are being sent into the four input modes of the 4SMS. The detection of single photons, one in output mode 3 and one in output mode 4 in the $\pm$-basis, heralds the generation of a two-photon Bell state in output modes 1 and 2 [see \cref{eq:herald4SMS}].}
    \label{fig:4SMS}
\end{figure}

Let us start with heralding the generation of Bell states. The 4-photon input state considered impinging on a 4SMS (see \cref{fig:4SMS}) is of the form 
\begin{align}
    \ket{\Psi_{\mu\eta, \text{in}}} & = \ket{1}_{\mu,1}\ket{1}_{\eta,2}\ket{1}_{\mu,3}\ket{1}_{\eta,4} \nonumber \\
    & \equiv \ket{\mu}_{1}\ket{\eta}_{2}\ket{\mu}_{3}\ket{\eta}_{4},
\end{align}
which consists of identical photons in the state $\ket{\mu}$ in modes 1 and 3, and identical photons in the state $\ket{\eta}$ in modes 2 and 4, respectively. Note that this state has been shown to generate 4-photon GHZ states when postselecting for 4-photon coincidences at the output of the 4SMS~\cite{Lim2005,Bhatti2023}.

From \cref{eq:Psi1-3,eq:Psi2-4} we know all possible output distributions which can result from two identical photons in input ports 1 and 3, and 2 and 4, respectively. It becomes clear that the detection of a single photon in output mode 3 (4), is necessarily linked to an identical photon in output mode 1 (2). At the same time, when detecting a single photon in each of the output modes 3 and 4, one knows that these two photons cannot be identical. This is exactly what allows for heralding the two-photon Bell states.

The complete output state with a single photon in output modes 3 and 4 each reads~\cite{Lim2005}
\begin{align}
    \ket{\Psi_{\mu\eta, \text{out}}} = \frac{1}{\sqrt{2}} \left( \ket{\eta}_{1}\ket{\mu}_{2}\ket{\eta}_{3}\ket{\mu}_{4}  - \ket{\mu}_{1}\ket{\eta}_{2}\ket{\mu}_{3}\ket{\eta}_{4} \right) ,
\end{align}
which is generated with a probability of $1/8$. This automatically means that detecting two single photons in modes 3 and 4 in the $\pm$-basis, i.e., $\ket{\pm}_{k} = (\ket{\mu}_{k} \pm \ket{\eta}_{k})/\sqrt{2}$ heralds a Bell state in modes 1 and 2. The exact form depends on the outcomes of the measurements:
\begin{align}
\label{eq:herald4SMS}
    \ket{+}_{3}\ket{+}_{4},\ket{-}_{3}\ket{-}_{4} &\rightarrow \frac{1}{\sqrt{2}} \left( \ket{\eta}_{1}\ket{\mu}_{2} - \ket{\mu}_{1}\ket{\eta}_{2} \right), \\
    \ket{+}_{3}\ket{-}_{4},\ket{-}_{3}\ket{+}_{4} &\rightarrow \frac{1}{\sqrt{2}} \left( \ket{\eta}_{1}\ket{\mu}_{2} + \ket{\mu}_{1}\ket{\eta}_{2} \right).
\end{align}

\begin{figure}
    \centering
    \includegraphics[width=0.85\linewidth]{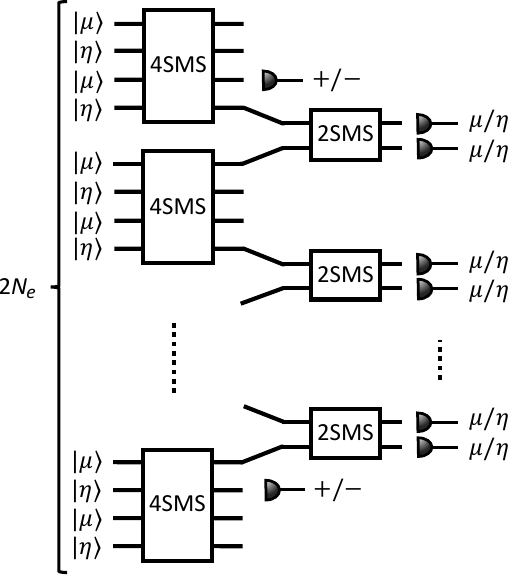}
    \caption{Setup to herald $N_{e}$-photon GHZ states using an array of four-port symmetric multiport splitters (4SMS) connected by two-port SMSs (2SMSs). $2N_{e}$ independent photons are sent into the $2N$ input modes of the multiports. The detection of single photons, one in output mode 3 of the first 4SMS and one in output mode 2 of the last 4SMS in the $\pm$-basis, together with the detection of two photons (one in the state $\ket{\mu}$ and one in $\ket{\eta}$) at every 2SMS, heralds the generation of an $N_{e}$-photon GHZ state in the remaining $N_{e}$ output modes.}
    \label{fig:4SMSGeneralized}
\end{figure}

Now, we extend this scheme to herald GHZ states of arbitrary photon numbers. For this, we connect multiple 4SMSs using two-port SMSs (2SMSs), i.e., 50/50 beam splitters (see \cref{fig:4SMSGeneralized}). At each 2SMS we detect exactly two orthogonal photons, i.e., one in the state $\ket{\mu}$ and one in the state $\ket{\eta}$, either in the same mode or in the two distinct modes. To guarantee that the heralding process works correctly, we have to make sure that the two photons have not been emitted from the same 4SMS. This can be done by additionally detecting a single photon in the $\pm$-basis in output mode 3 (2) of the first (last) 4SMS. In particular, if we detect a photon in output mode 3 (2) of the first (last) 4SMS, we immediately know that there has to be an identical photon in output mode 1 (4) of the same 4SMS. Therefore, the two orthogonal photons detected at the first (last) beam splitter cannot both have come from the same 4SMS. For only two connected 4SMSs, this directly proofs that the four undetected photons are projected into a GHZ state, i.e.,
\begin{align}
    \ket{\Psi_{\text{GHZ}}} = & \frac{1}{\sqrt{2}} \left( \ket{\eta}_{1,1}\ket{\mu}_{1,2}\ket{\eta}_{2,3}\ket{\mu}_{2,4} \right. \nonumber \\
    & \left. \pm \ket{\mu}_{1,1}\ket{\eta}_{1,2}\ket{\mu}_{2,3}\ket{\eta}_{2,4} \right),
\end{align}
where the double index $i,j$ denotes the 4SMS $i$ and the output mode $j$. The phase depends on the photons detected in the $\pm$-basis and the detection pattern at the beam splitter.

Following this logic through the chain of 4SMSs (see \cref{fig:4SMSGeneralized}) proves that GHZ states of an arbitrary even number $N_{e}$ can be heralded. The success probability is given by
\begin{align}
    p_{N_{e}} = \left( \frac{1}{8} \right)^{\frac{N_{e}}{2}} \left( \frac{1}{2} \right)^{\frac{N_{e}}{2}-1} = \left( \frac{1}{2} \right)^{2N_{e}-1} ,
\end{align}
where each of the $N_{e}/2$ 4SMSs succeeds with a probability of $1/8$ and two neighboring 4SMSs can be connected successfully with a probability of $1/2$.

To generate GHZ states of odd numbers $N_{o}$, one can reduce the number of modes by one by additionally detecting photons in the $\pm$-basis in the output mode 4 (1) of the last (first) 4SMS. Accepting every two-photon event in the two modes 2 and 4 (1 and 3) changes the success probability of this particular 4SMS from $1/8$ to $1/4$ [see \cref{eq:Psi1-3,eq:Psi2-4}]. The total success probability for odd $N_{o}$ derives to
\begin{align}
    p_{N_{o}} = \frac{1}{4} \left( \frac{1}{8} \right)^{\frac{N_{o}+1}{2}-1} \left( \frac{1}{2} \right)^{\frac{N_{o}+1}{2}-1}  =  \left( \frac{1}{2} \right)^{2N_{o}} .
\end{align}

Note that the presented scheme and the derived scaling of the success probability are similar to other schemes known from the literature~\cite{Zhang2008,Gimeno-Segovia2015Thesis,Bartolucci2021,Shimizu2024}. In particular, for $N_{e}$ the scaling is identical to Refs.~\cite{Gimeno-Segovia2015Thesis} and \cite{Bartolucci2021}, which use polarization and path encoding, respectively. Additionally, for $N_{e}=4$, by using feedforward and entanglement distillation the success probability in those schemes can even be increased further~\cite{Zhang2008,Joo2007,Bartolucci2021}, which is not possible here. However, our scheme has the advantage that it works for arbitrary degrees of freedom.

\section{Heralding higher-dimensional Bell states using 3-port splitters}
\label{sec:Bell3SMS}

In this section, we describe a simple scheme for heralding higher-dimensional Bell states of arbitrary dimension $d$. The scheme requires $3d$ photons, $d$ 3SMSs, one $d$-port SMS (dSMS) and photon-number resolution. Further, we discuss that with the appropriate input states and measurements, this scheme could also be implemented using a single 3SMS only.

Let us start with three identical photons injected into the three input ports of a 3SMS [see \cref{eq:PsiIn,eq:tritter}]. From the output state [see \cref{eq:PsiOutTritter}]
one can directly see, that detecting a single photon in output mode 1 heralds a single photon in output modes 2 and 3 each. This happens with a probability of $1/3$. Detecting three photons in output mode 1 heralds vacuum in the other two modes. This happens with a probability of $2/9$.

\begin{figure}
    \centering
    \includegraphics[width=0.85\linewidth]{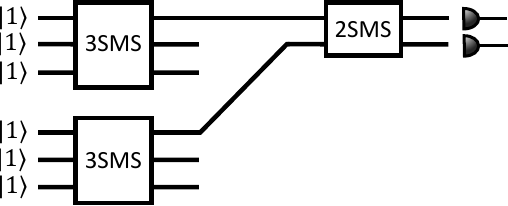}
    \caption{Setup to herald a two-photon qubit Bell state using two three-port symmetric multiport splitter (3SMS) connected by a single two-port SMS (2SMS) (see \cref{sec:Bell3SMS}). Detecting four photons at the 2SMS directly heralds the generation of a path-encoded Bell state in the remaining modes [see \cref{eq:BellQubit}].}
    \label{fig:Setup3SMS_Two}
\end{figure}

Now we connect two identical 3SMSs using a 2SMS (see \cref{fig:Setup3SMS_Two}) and detect in total four photons at the 2SMS. Such a detection event can only result from $3+1$ or $1+3$ photons, i.e., three photons emitted by the first 3SMS and one photon by the second 3SMS or the other way around. Each of these events has a probability of $2/27$. Thus, this scheme can be used to herald the following two-photon Bell state in path encoding with a probability of $p_{2\text{Bell}}=4/27$:
\begin{align}
\label{eq:BellQubit}
    & \ket{\Phi_{2\text{Bell}}} \nonumber \\
    & =  \frac{1}{\sqrt{2}} \left( \ket{1}_{1,2}\ket{1}_{1,3}\ket{0}_{2,2}\ket{0}_{2,3} \pm \ket{0}_{1,2}\ket{0}_{1,3}\ket{1}_{2,2}\ket{1}_{2,3} \right) \nonumber \\
    & \equiv \frac{1}{\sqrt{2}} \left( \ket{1,1}_{1}\ket{0,0}_{2} \pm \ket{0,0}_{1}\ket{1,1}_{2} \right),
\end{align}
where the phase $\pm$ depends on the four-photon detection pattern. In line 2 we defined $\ket{1/0}_{i,j}\ket{1/0}_{i,j+1} \equiv \ket{1/0,1/0}_{i}$, with $i$ denoting the number of the 3SMS, and $j$ and $j+1$ denoting the numbers of the output modes. 
Note that, since the state $\ket{3}_{1,1}\ket{1}_{2,1} + \ket{1}_{1,1}\ket{3}_{2,1}$ in the heralding modes of the two 3SMSs results in a four-photon $N00N$ state using a 2SMS~\cite{Matthews2011}, this setup can also be used to herald the four-photon $N00N$ state. For this, one would have to detect the two-photon Bell state given in \cref{eq:BellQubit} instead.

\begin{figure}
    \centering
    \includegraphics[width=0.85\linewidth]{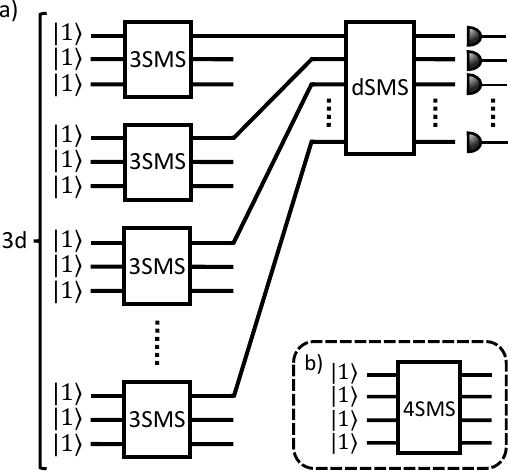}
    \caption{a) Setup to herald $d$-dimensional two-photon Bell states using an array of 3-port symmetric multiport splitters (3SMSs) connected by a single $d$-port SMS (dSMS) (see \cref{sec:Bell3SMS}). $3d$ independent, identical photons are sent into the $3d$ input modes of the multiports. The detection of $3d-2$ photons at the output modes of the dSMS heralds the generation of a $d$-dimensional two-photon Bell state in the remaining $2d$ output modes, two per 3SMS. b) The same scheme can be used to herald the generation of three-photon $d$-dimensional GHZ states when exchanging the 3SMSs with 4-port SMSs (4SMSs) and detecting $4d-3$ photons at the output modes of the dSMS (see \cref{sec:GHZ4SMS}).
    }
    \label{fig:Setup3SMS}
\end{figure}

Next, let us discuss how to extend this scheme to herald $d$-dimensional Bell states. For this, we insert $3d$ photons into the input modes of $d$ 3SMSs, combine one output mode of each 3SMS in a dSMS, and detect $3d-2$ photons at the dSMS [see \cref{fig:Setup3SMS} a)]. Such a detection event can only follow from exactly one 3SMS emitting a single photon and all other 3SMSs emitting three photons each. Thus, it directly heralds the generation of a $d$-dimensional Bell state of the form
\begin{align}
\label{eq:higherdimensionalBellstate}
    \ket{\Phi_{d\text{Bell}}} & =  \frac{1}{\sqrt{d}} \left( e^{i \varphi_1} \ket{1,1}_{1}\ket{0,0}_{2}\ldots\ket{0,0}_{d} \right. \nonumber \\
    & \phantom{={}} +  e^{i \varphi_2} \ket{0,0}_{1}\ket{1,1}_{2}\ket{0,0}_{3}\ldots\ket{0,0}_{d} \nonumber \\
    &\phantom{={}} \left. + \cdots + e^{i \varphi_d} \ket{0,0}_{1}\ldots\ket{0,0}_{d-1}\ket{1,1}_{d} \right) ,
\end{align}
where each term of the Bell state has identical weight due to the symmetry of the dSMS and the possible input states (see \cref{App:3SMS} for mathematical proof). However, the phases $\varphi_i$, with $i=1,2,\ldots,d$ depend on the exact $3d-2$-photon detection patterns.

The success probability is given by
\begin{align}
\label{eq:probabilityBell}
    p_{d\text{Bell}} = d \times \frac{1}{3} \left( \frac{2}{9} \right)^{d-1} = \frac{d \times 2^{d-1}}{3^{2d -1}} ,
\end{align}
with $1/3$ ($2/9$) being the probability of a 3SMS to produce the state $\ket{1,1,1}$ ($\ket{3,0,0}$), and $d$ being the number of permutations. For example, for $d=3$ we find $p_{3\text{Bell}} = 4/81$. Comparing \cref{eq:probabilityBell} to other heralding schemes from the literature~\cite{Paesani2021,Chin2024}, we find that our result demonstrates improved scaling but needs more auxiliary photons.
In \cref{App:Bell} we investigate an additional scheme to generate higher-dimensional Bell states which uses 2SMSs as basic elements. Although this scheme demonstrates an even higher success probability [see \cref{eq:probability_dBell_ff} and \cref{fig:Prob_Bell}], it also needs more photons, i.e., $4d$ instead of $3d$.

In the case of $d=3$, detecting five instead of seven photons
directly heralds an inverted three-dimensional four-photon Bell state of the form
\begin{align}
\label{eq:BellStateQuditInverted}
    \ket{\Phi_{3\overline{\text{Bell}}}} & = \frac{1}{\sqrt{3}} \left( e^{i \varphi_1} \ket{0,0}_{1}\ket{1,1}_{2}\ket{1,1}_{3} \right. \nonumber \\
    & \phantom{={}}  + e^{i \varphi_2} \ket{1,1}_{1}\ket{0,0}_{2}\ket{1,1}_{3} \nonumber \\
    & \phantom{={}} \left. + e^{i \varphi_3} \ket{1,1}_{1}\ket{1,1}_{2}\ket{0,0}_{3} \right) .
\end{align}
The probability for this to occur is $p_{3\overline{\text{Bell}}}=3\times 2/9\times(1/3)^{2} = 2/27$. Heralding both, $\ket{\Phi_{3\text{Bell}}}$ and $\ket{\Phi_{3\overline{\text{Bell}}}}$, gives a total success probability of $p_{\sum 3\text{Bell}}=10/81 \approx 0.12$. In \cref{tab:Bell} we show an overview over the different results for $d=3$.
Note that, the inverted Bell state generation does not work for $d>3$, since the increasing number of photon output distributions affects the heralding.

\renewcommand{\arraystretch}{1.75}
\setlength{\tabcolsep}{10pt}
\begin{table}
    \centering
    \begin{tabular}{c|ccc}
    Bell state     &  $\ket{\Phi_{3\text{Bell}}}$ & $\ket{\Phi_{3\overline{\text{Bell}}}}$ & both \\
    \hline
    Probability     & $\frac{4}{81}$  & $\frac{2}{27}$  & $\frac{10}{81}$ \\
    \end{tabular}
    \caption{Overview over the different heralded Bell states and their success probabilities for $d=3$. $\ket{\Phi_{3\text{Bell}}}$ denotes the two-photon Bell state in path-encoding [see \cref{eq:higherdimensionalBellstate,eq:probabilityBell}], $\ket{\Phi_{3\overline{\text{Bell}}}}$ denotes the inverted, i.e., four-photon Bell state in path-encoding [see \cref{eq:BellStateQuditInverted}].}
    \label{tab:Bell}
\end{table}

Finally, we argue that the whole setup could in principle be implemented using a single 3SMS for arbitrary photonic degrees of freedom capable of implementing a qudit system. For this, one would have to insert $d$ photons per every input mode of the 3SMS, to be exact one photon per qudit state. Additionally, one would have to detect $3d-2$ photons in one of the output modes indistinguishably in one of the orthogonal bases.
In the case of $d=2$, this can easily be done using polarization encoding and employing polarizing beam splitters to combine and to detect photons. In the case of larger $d$ one has to think of more complicated setups. For example, using orbital angular momentum, one could use dSMSs and Dove-prisms to combine the photons and perform the required measurements~\cite{Kysela2020}. Preferably, using time-bin encoding, one would inject $3d-2$ photons for every time bin, and use dSMSs and optical switches for implementing the measurements~\cite{Zheng2022}.

\section{Heralding higher-dimensional GHZ states using 4-port splitters}
\label{sec:GHZ4SMS}

We now show that the scheme presented in \cref{sec:Bell3SMS} can be adjusted to herald the generation of higher-dimensional three-party GHZ states. For this, one simply has to exchange the 3SMSs by 4SMSs, which increases the total number of input modes and photons to $4d$, respectively.

Let us start with a single 4SMS. Sending in four independent, identical photons, the output is given by \cref{eq:PsiOutQuitter}.
From this, we see that detecting a single photon with a probability of $1/4$ in output mode 1 heralds the state
\begin{align}
    \ket{1}_{1} \rightarrow \frac{1}{2\sqrt{2}} \left( \ket{2}_{2}\ket{1}_{3}\ket{0}_{4}  + \ket{0}_{2}\ket{1}_{3}\ket{2}_{4} \right),
\end{align}
which can deterministically be transformed into the wanted state $\ket{1}_{2}\ket{1}_{3}\ket{1}_{4}$. On the other hand, detecting 4 photons in output mode 1 with a probability of $3/32$, heralds the state $\ket{0}_{2}\ket{0}_{3}\ket{0}_{4}$.

Now, connecting two 4SMSs using a 2SMS (see \cref{fig:Setup3SMS}) and detecting five photons heralds a three-photon path-encoded GHZ state
\begin{align}
    \ket{\Phi_{2\text{GHZ}}} =  \frac{1}{\sqrt{2}} \left( \ket{1,1,1}_{1}\ket{0,0,0}_{2} \pm \ket{0,0,0}_{1}\ket{1,1,1}_{2} \right) ,
\end{align}
where again the phase $\pm$ depends on the photon detection pattern. This state can be heralded with a probability of $p_{2\text{GHZ}}=2\times 1/4 \times 3/32 = 3/64$.

Connecting $d$ 4SMSs by a single dSMS (see \cref{fig:Setup3SMS}) and detecting $4d-3$ photons at the dSMS heralds the $d$-dimensional three-party GHZ state
\begin{align}
\label{eq:higherdimensionalGHZstate}
    &\ket{\Phi_{d\text{GHZ}}} \nonumber \\
    & =  \frac{1}{\sqrt{d}} \left( e^{i \varphi_1} \ket{1,1,1}_{1}\ket{0,0,0}_{2}\ldots\ket{0,0,0}_{d} \right. \nonumber \\
    & \phantom{={}} +  e^{i \varphi_2} \ket{0,0,0}_{1}\ket{1,1,1}_{2}\ket{0,0,0}_{3}\ldots\ket{0,0,0}_{d} \nonumber \\
    & \phantom{={}} \left. + \cdots + e^{i \varphi_d} \ket{0,0,0}_{1}\ldots\ket{0,0,0}_{d-1}\ket{1,1,1}_{d} \right) ,
\end{align}
where, again, each term of the GHZ state has identical weight due to the symmetry of the dSMS and the possible input states (see \cref{App:3SMS} for mathematical proof). The phases $\varphi_i$, with $i=1,2,\ldots,d$ depend on the exact $4d-3$-photon detection patterns.

The success probability derives to
\begin{align}
\label{eq:probabilityGHZ}
    p_{d\text{GHZ}} = d \times \frac{1}{4} \left( \frac{3}{32} \right)^{d-1} = \frac{d \times 3^{d-1}}{2^{5d -3}} ,
\end{align}
with $1/4$ ($3/32$) being the probability of a 4SMS to produce the state $\ket{1,1,1,1}$ ($\ket{4,0,0,0}$), and $d$ being the number of permutations. For example, for $d=3$, this gives $p_{3\text{GHZ}}\approx6.6\times 10^{-3}$. Comparing this result to other heralding schemes from the literature~\cite{Paesani2021,Chin2024}, we find that our scheme has an improved scaling. Further, it is resource efficient; it needs fewer modes and photons compared to~\cite{Paesani2021}, less optical elements compared to~\cite{Chin2024}.

Similar to \cref{sec:Bell3SMS}, we argue that the whole setup could in principle be implemented using a single 4SMS assuming that the necessary input states, i.e., one photon per qudit state per input mode, can be generated, and the necessary measurements can be implemented. Again, time-bin encoding might be most suited for such an implementation.

\section{Increasing the success probability using deterministic photon subtraction}
\label{sec:PhotonSubtraction}

\begin{figure}
    \centering
    \includegraphics[width=0.85\linewidth]{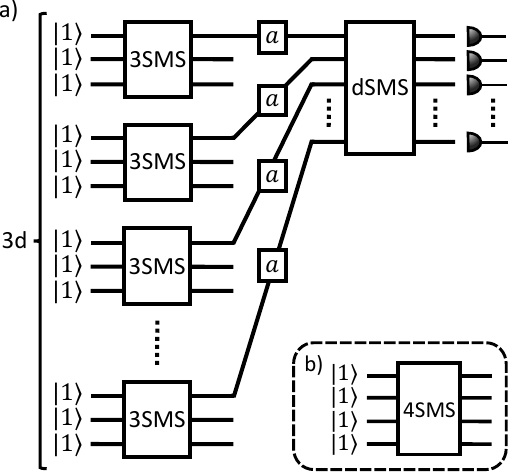}
    \caption{a) Setup from \cref{sec:Bell3SMS} to herald $d$-dimensional two-photon Bell states with additional photon subtraction to increase the success probability (see \cref{sec:PhotonSubtraction}). $3d$ independent, identical photons are sent into the $3d$ input modes of $d$ 3-port symmetric multiport splitters (3SMSs).
    In each heralding mode deterministic photon subtraction $a$ subtracts and detects a single photon, i.e., in total $d$ photons are subtracted. The detection of additional $2d-2$ photons at the output modes of the $d$-port SMS (dSMS) heralds the generation of a $d$-dimensional two-photon Bell state in the remaining $2d$ output modes, two per 3SMS. b) The same scheme including photon subtraction can be used to herald the generation of three-photon $d$-dimensional GHZ states when exchanging the 3SMSs with 4-port SMSs (4SMSs). Photon subtractions in the $d$ heralding modes subtract and detect $d$ photons. Additionally, detecting $3d-3$ photons at the output modes of the dSMS heralds the $d$-dimensional GHZ state (see \cref{sec:PhotonSubtraction}).
    }
    \label{fig:Setup3SMSSubtraction}
\end{figure}

In Ref.~\cite{Bartolucci2021} the authors discuss a method, the so-called bleeding, in which they use deterministic photon subtractions to increase the success probability of heralding the generation of qubit Bell and GHZ states. In this section, we show that the same method also increases the success probabilities of the schemes described in~\cref{sec:Bell3SMS,sec:GHZ4SMS}.

For the following discussion, let us simply assume that the bleeding operation subtracts and detects a single photon from the heralding mode of each 3SMS (4SMS) (see \cref{fig:Setup3SMSSubtraction}); for a detailed discussion of the method including losses see~\cite{Bartolucci2021}. This means that if and only if the heralding mode of a 3SMS (4SMS) is occupied, the subtraction is successful and the respective output state is accepted and stored.
If no photon is detected, we assume that the heralding mode is empty and this part, i.e., 3SMS (4SMS), has to be repeated. In this way, all 3SMSs (4SMSs) can be repeated until successful in parallel. This reduces the number of overall tries compared to the case when all 3SMSs (4SMSs) have to be successful coincidentally. Finally, we use $d$ subtracted output states for heralding qudit Bell (GHZ) states.

\begin{figure}[t]
    \centering
    \includegraphics[width=1.0\linewidth]{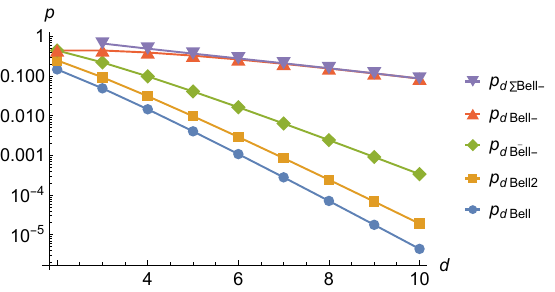}
    \caption{Plot of the different success probabilities for heralding a $d$-dimensional Bell state without and with photon subtraction. Here, the plotted probabilities do not take into account the probabilities for preparing the subtracted states but only the final heralding. The Bell state generation success probabilities from schemes without photon subtraction, i.e., $p_{d\text{Bell}}$ and $p_{d\text{Bell}2}$, demonstrate the lowest values [see \cref{eq:probabilityBell,eq:probability_dBell_ff}]. In comparison, the Bell state generation success probabilities with deterministic photon subtraction $p_{d\text{Bell}-}$, $p_{d\overline{\text{Bell}}}$, and $p_{d\sum \text{Bell}-}$ demonstrate significantly higher values [see \cref{eq:probabilityBellSub,eq:probabilityBellSubInvert,eq:probabilityBellSubTotal}].}
    \label{fig:Prob_Bell}
\end{figure}

In the case of heralded qudit Bell states (see \cref{sec:Bell3SMS}), the photon subtracted output state of the 3SMS takes the form [see \cref{eq:PsiOutTritter}]
\begin{align}
\label{eq:PsiOutTritterSub}
    \ket{\Psi_{3-}} = & a_{1} \ket{\Psi_{3,\text{out}}} = \frac{\sqrt{2}}{\sqrt{3}}  \ket{2,0,0}  - \frac{1}{\sqrt{3}} \ket{0,1,1}.
\end{align}
Now, to herald a $d$-dimensional Bell state with two photons one needs to detect $2d-2$ additional photons at the connecting dSMS [see \cref{fig:Setup3SMSSubtraction} a)]. This gives the adjusted success probability [see \cref{eq:probabilityBell}]
\begin{align}
\label{eq:probabilityBellSub}
    p_{d\text{Bell}-} = d \times \frac{1}{3} \left( \frac{2}{3} \right)^{d-1} = \frac{d\times 2^{d-1}}{3^{d}} .
\end{align}
Furthermore, we see that using photon subtraction one can herald the inverted $d$-dimensional Bell state with $2d-2$ photons by detecting two photons for arbitrary $d$. The success probability is given by
\begin{align}
\label{eq:probabilityBellSubInvert}
    p_{d\overline{\text{Bell}}-} = d \times \frac{2}{3} \left( \frac{1}{3} \right)^{d-1}  = \frac{d \times 2}{3^{d}} .
\end{align}
For $d\geq3$, allowing both possible Bell state generations, we obtain the total success probability
\begin{align}
\label{eq:probabilityBellSubTotal}
   p_{d\sum\text{Bell}-} =  p_{d\text{Bell}-} + p_{d\overline{\text{Bell}}-} =  \frac{d \times (2+2^{d-1})}{3^{d}} ,
\end{align}
which, for $d=3$, becomes $2/3 \approx 0.66$. To show the scaling of the different Bell state success probabilities, we plot them in \cref{fig:Prob_Bell}, including \cref{eq:probabilityBell,eq:probability_dBell_ff} without photon subtraction. One can see that the probabilities with photon subtraction demonstrate a highly improved scaling, especially $p_{d\text{Bell}-}$. Although $p_{d\overline{\text{Bell}}-}$ demonstrates lower values, this state generation only requires the additional detection of two photons following the $d$ photon subtractions.

\begin{figure}
    \centering
    \includegraphics[width=1.0\linewidth]{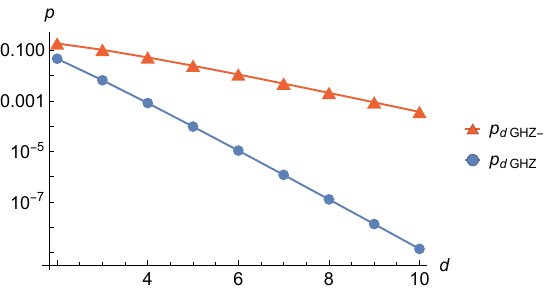}
    \caption{Plot of the different success probabilities for heralding a $d$-dimensional GHZ state without and with photon subtraction. Here, the plotted probabilities do not take into account the probabilities for preparing the subtracted states but only the final heralding. The figure compares the GHZ state generation success probabilities without ($p_{d\text{GHZ}}$) and with ($p_{d\text{GHZ}-}$) photon subtraction [see \cref{eq:probabilityGHZ,eq:probabilityGHZsub}].}
    \label{fig:Prob_GHZ}
\end{figure}

In the case of the heralded qudit GHZ states (see \cref{sec:GHZ4SMS}), the photon subtracted output states of every 4SMS takes the form [see \cref{eq:PsiOutQuitter}]
\begin{align}
\label{eq:PsiOutQuitterSub}
    \ket{\Psi_{4-}} & =  a_{1} \ket{\Psi_{4,\text{out}}} \nonumber \\
    & = \frac{\sqrt{3}}{2\sqrt{2}} \ket{3,0,0,0} - \frac{1}{2}  \ket{1,1,0,1} \nonumber \\
    & \phantom{={}} + \frac{1}{2\sqrt{2}} \left( \ket{0,2,1,0} + \ket{0,0,1,2}  -  \ket{1,0,2,0} \right) .
\end{align}
To herald the $d$-dimensional GHZ state one needs to detect $3d-3$ additional photons at the connecting dSMS [see \cref{fig:Setup3SMSSubtraction} b)]. The success probability calculates to
\begin{align}
\label{eq:probabilityGHZsub}
    p_{d\text{GHZ}-} = d \times \frac{1}{4} \left( \frac{3}{8} \right)^{d-1} = \frac{d \times 3^{d-1}}{2^{3d -1}} .
\end{align}
We plot the success probability and compare it to the probability without photon subtraction, i.e., \cref{eq:probabilityGHZ}, in \cref{fig:Prob_GHZ}. Note that, in contrast to the Bell state generation one cannot herald the generation of an inverted GHZ state with $3d-3$ photons due to the additional terms in \cref{eq:PsiOutQuitterSub}.

\begin{figure}
    \centering
    \includegraphics[width=1.0\linewidth]{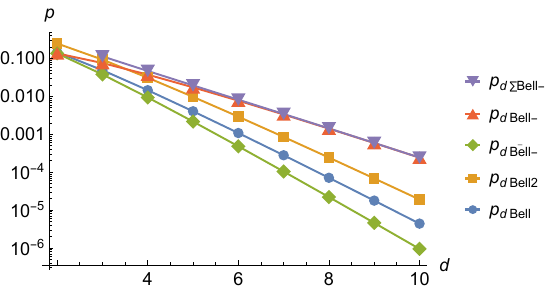}
    \caption{Plot of the different success probabilities for heralding a $d$-dimensional Bell state without and with photon subtraction. Here, it is assumed that all photon subtractions have to be successful within the same try. Compared to \cref{fig:Prob_Bell} each probability with photon subtraction, i.e., $p'_{d\text{Bell}-}$, $p_{d'\overline{\text{Bell}}}$, and $p'_{d\sum \text{Bell}-}$, has been multiplied by a factor of $(5/9)^{d}$. Especially for larger $d$, $p'_{d\text{Bell}-}$ and $p'_{d\sum \text{Bell}-}$ have an advantage over $p_{d\text{Bell}}$.}
    \label{fig:Prob_Bellprime}
\end{figure}

\begin{figure}
    \centering
    \includegraphics[width=1.0\linewidth]{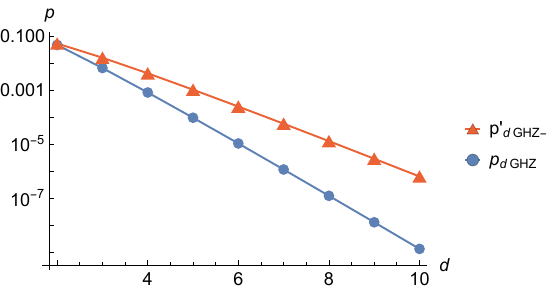}
    \caption{Plot of the different success probabilities for heralding a $d$-dimensional GHZ state without and with photon subtraction. Here, it is assumed that all photon subtractions have to be successful within the same try. Compared to \cref{fig:Prob_GHZ} the probability with photon subtraction, i.e., $p'_{d\text{GHZ}-}$, has been multiplied by a factor of $(17/32)^{d}$. Especially for larger $d$, $p'_{d\text{GHZ}-}$ has an advantage over $p_{d\text{GHZ}}$.}
    \label{fig:Prob_GHZprime}
\end{figure}

Finally, we show that the deterministic photon subtraction is of advantage even when assuming that all $d$ photon subtractions and the final heralding have to be successful in one try.
The probability that the photon subtraction works is given by the probability that the heralding mode of the 3SMS (4SMS) is not empty. From \cref{eq:PsiOutTritter} [\cref{eq:PsiOutQuitter}] we find that this probability is given by $5/9$ ($17/32$). 
Therefore, to include the probability that all photon subtractions give a positive result, i.e., none of the heralding modes is empty, \cref{eq:probabilityBellSub,eq:probabilityBellSubInvert,eq:probabilityBellSubTotal} [\cref{eq:probabilityGHZsub}] have to be multiplied by a factor of $(5/9)^{d}$ [$(17/32)^{d}$]. We plot the multiplied success probabilities for the Bell (GHZ) state generation in \cref{fig:Prob_Bellprime} (\cref{fig:Prob_GHZprime}).

From \cref{fig:Prob_Bellprime,fig:Prob_GHZprime} one can see that even when all 3SMSs (4SMSs) have to be successful in the same try deterministic photon subtraction boosts the overall success probability. This is especially true for increasing $d$. The reason for this is that photon subtraction changes the weights between the different terms in the subtracted state~\cite{Barnett2018}.

\section{Conclusion}
\label{sec:Conclusion}

In this paper, we discussed easily implementable schemes to herald qubit GHZ states, higher-dimensional Bell states and higher-dimensional three-party GHZ states. The presented schemes demonstrate high success probabilities and need assessable numbers of auxiliary modes, photons, and optical elements. All presented schemes can be adjusted to work for arbitrary internal degrees of freedom. In the final part of the paper we have discussed how deterministic photon subtraction can be used to increase the success probabilities for the schemes heralding qudit Bell and GHZ states even further.

Due to the increasing number of possible output distributions for dSMSs with $d>4$, no generalization of the presented schemes has been found. However, by heralding the generation of Bell states and three-party GHZ states of arbitrary dimensions one has the building blocks for fusing arbitrary graph states, which are required for multiparty quantum communication schemes and high-dimensional measurement-based quantum computing~\cite{Murta2020,Gimeno-Segovia2015,Paesani2021}.

The aim of this work was to herald entangled quantum states with zero or one photons in each mode. However, heralding entangled quantum states with multiple photons in the same output modes could allow for generating logical quantum states suitable for quantum error-correction codes against photon losses~\cite{Bergmann2016}. Furthermore, it will be interesting to investigate if the presented schemes can be adjusted to directly herald different types of states, e.g., the generation of cluster states~\cite{Thomas2024}, or multiphoton Dicke states~\cite{Thiel2007}. One way of increasing the number of different output states but also the success probabilities could be to apply photon subtraction multiple times, either in the same or in different output modes.

While finishing the manuscript, we became aware of Ref.~\cite{Chin2024}. This work also discusses the generation of qudit GHZ states using linear optics, however, for arbitrary dimensions. The presented scheme is different from our scheme in design and in scaling. It requires less input photons but more and more complicated optical elements.

\begin{acknowledgments}
We thank Simone E. D’Aurelio for comments on the manuscript. We acknowledge support from the Carl Zeiss Foundation, the Center for Integrated Quantum Science and Technology (IQ$^\text{ST}$), the Federal Ministry of Education and Research (BMBF, projects SiSiQ and PhotonQ), the Federal Ministry for Economic Affairs and Climate Action (BMWK, project PlanQK), and the Competence Center Quantum Computing Baden-W\"urttemberg (funded by the Ministerium f\"ur Wirtschaft, Arbeit und Tourismus Baden-W\"urttemberg, project QORA). D.B. was
partially supported by the JST Moonshot R\&D program under Grant JPMJMS226C.
\end{acknowledgments}

\appendix

\section{Heralding higher-dimensional Bell states using 2-port splitters}
\label{App:Bell}

In Ref.~\cite{Bartolucci2021} a scheme using eight photons and feedforward has been introduced to herald the generation of qubit Bell states with a probability $1/4$. Here, we show that this scheme can be extended to herald the generation of qudit Bell states.

\begin{figure}
    \centering
    \includegraphics[width=0.85\linewidth]{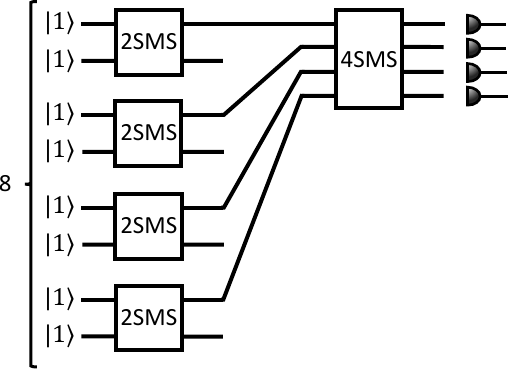}
    \caption{Setup to herald two-photon qubit Bell states using an array of balanced beam splitters (2SMSs). Four 2SMSs are connected by a single four-port symmetric multiport splitter (4SMS). Detecting six photons at the outputs of the 4SMS generates a superposition of two photons in one of the remaining modes [see \cref{eq:2Bellff}]. This state can be transformed to a path-encoded qubit Bell state using additional 2SMSs and phases, depending on the exact detection pattern at the 4SMS. The figure has been adapted from Ref.~\cite{Bartolucci2021}.}
    \label{fig:2SMSQudit}
\end{figure}

In the basic scheme (see \cref{fig:2SMSQudit}) eight single photons are inserted into the eight input ports of four 2SMSs~\cite{Bartolucci2021}. Each 2SMS generates the state
\begin{align}
    \ket{\Psi_{2}} = \frac{1}{\sqrt{2}} \left( \ket{2,0} - \ket{0,2} \right) .
\end{align}
Connecting the four 2SMSs using a 4SMS and detecting six photons directly heralds the generation of the state
\begin{align}
\label{eq:2Bellff}
    & \ket{\Psi_{\text{2Bell2}}} \nonumber \\
    & = \frac{1}{2} \left( e^{\varphi_{1}}\ket{2}_{1}\ket{0}_{2}\ket{0}_{3}\ket{0}_{4} + e^{\varphi_{2}}\ket{0}_{1}\ket{2}_{2}\ket{0}_{3}\ket{0}_{4} \right. \nonumber \\
    & \phantom{={}} \left. +\ e^{\varphi_{3}}\ket{0}_{1}\ket{0}_{2}\ket{2}_{3}\ket{0}_{4} + e^{\varphi_{4}}\ket{0}_{1}\ket{0}_{2}\ket{0}_{3}\ket{2}_{4} \right) ,
\end{align}
where the phases depend on the detection pattern at the 4SMS. By using appropriate phases and additional 2SMSs between the output modes, this state can be transformed into a path-encoded two-photon Bell state. The probability for this is $p_{2\text{Bell2}}=4\times (1/2)^{4}= 1/4$~\cite{Bartolucci2021}.

\begin{figure}
    \centering
    \includegraphics[width=0.85\linewidth]{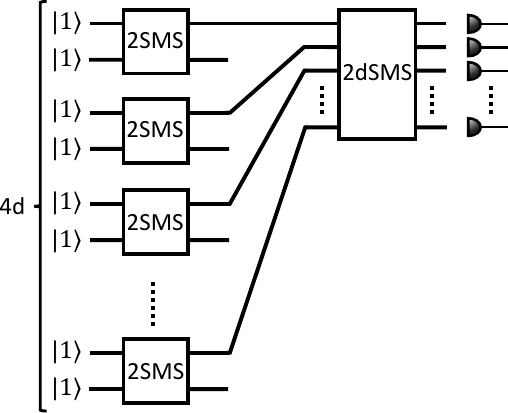}
    \caption{Setup to herald two-photon qudit Bell states using an array of balanced beam splitters (2SMSs). $2d$ 2SMSs with in total $4d$ input modes are connected by a single $2d$-port symmetric multiport splitter (2dSMS). Detecting $4d-2$ photons at the outputs of the 2dSMS generates a superposition of two photons in one of the remaining modes [see \cref{eq:dBellff}]. This state can be transformed to a path-encoded qudit Bell state using additional 2SMSs and phases, depending on the exact detection pattern at the 2dSMS.}
    \label{fig:2SMSQudit2}
\end{figure}

Similar to the schemes presented in \cref{sec:Bell3SMS,sec:Bell4portSMS}, this scheme can be extended to herald $d$-dimensional two-photon qudit Bell states. For this, we connect $2d$ 2SMSs using a single 2dSMS and detect $4d-2$ photons at the 2dSMS (see \cref{fig:2SMSQudit2}). The heralded state in the remaining modes takes the form
\begin{align}
\label{eq:dBellff}
    \ket{\Psi_{d\text{Bell2}}}  &= \frac{1}{\sqrt{2d}} \left( e^{\varphi_{1}}\ket{2}_{1}\ket{0}_{2}\ldots\ket{0}_{2d} \right. \nonumber \\
    & \phantom{={}} \left. + \dots + e^{\varphi_{2d}}\ket{0}_{1}\ldots\ket{0}_{2d-1}\ket{2}_{2d} \right) ,
\end{align}
where, again, the phases depend on the $4d-2$-photon detection pattern at the 2dSMS. All terms have the same weight due to the symmetry of the 2dSMS and the possible input states (see \cref{App:3SMS} for mathematical proof). This state can be transformed into a path-encode two-photon Bell state. The success probability to herald this state is given by
\begin{align}
\label{eq:probability_dBell_ff}
    p_{d\text{Bell2}}=2d \times \left(\frac{1}{2}\right)^{2d} = \frac{d}{2^{2d-1}}.
\end{align}
We plot this probability in \cref{fig:Prob_Bell,fig:Prob_Bellprime} to compare it to the success probability of the schemes derived in \cref{sec:Bell3SMS,sec:PhotonSubtraction} using 3SMSs without and with deterministic photon subtraction. One can see that the scheme using 2SMSs demonstrates a better scaling. However, it needs more photons, i.e., $4d$ compared to $3d$. Also, deterministic photon subtraction cannot be applied to increase the success probability.

\section{Weights of higher-dimensional Bell (GHZ) state}
\label{App:3SMS}

Here, we prove that all terms of the heralded higher-dimensional states given in \cref{eq:higherdimensionalBellstate,eq:higherdimensionalGHZstate,eq:dBellff} have the same weight, respectively. We assume that we detect $m$ photons at the output of the connecting $D$-port SMS in an arbitrary output distribution (for definitions of $m$ and $D$ see \cref{tab:Values}). The set of output modes belonging to the distribution is given by $\sigma = \{ k_1, k_2, \ldots, k_m \}$, where $k_i \in \{ 1,2,\ldots, D \}$ denotes the output mode of the $i$th photon.

\begin{table}[b]
    \centering
    \begin{tabular}{c|cccc}
         & $m$  & $D$ & $m_1$ & $m_2$ \\ \hline
        Bell [\cref{eq:higherdimensionalBellstate}] & $3d-2$ & $d$ & 1 & 3 \\
        GHZ [\cref{eq:higherdimensionalGHZstate}] & $4d-3$ & $d$ & 1 & 4 \\
        Bell [\cref{eq:dBellff}] & $4d-2$ & $2d$ & 0 & 2 \\
    \end{tabular}
    \caption{Definitions of photon numbers and number of ports needed in the schemes generating $d$-dimensional Bell and GHZ states. $m$ denotes the total number of photons, which has to be detected at the connecting $D$-port symmetric multiport splitter (DSMS). $m_1$ and $m_2$ denote the different numbers of photons, which can enter the DSMS through a single port.}
    \label{tab:Values}
\end{table}

Now, we know that there are only $D$ different possible input distributions, i.e., one input mode has $m_1$ photons and all other input modes have $m_2$ photons each (for definitions of $m_1$ and $m_2$ see \cref{tab:Values}). We randomly choose one distribution and prove that this and the next distribution according to cyclic permutation lead to the output $\sigma$ with the same probability.

The chosen input is described by its set of input modes $\gamma = \{ l_1, l_2, \ldots, l_m \}$, and the transition probability can be written as~\cite{Bhatti2023}
\begin{align}
	\tilde{p}_{\sigma}  = \frac{1}{N^{m}} \big| \mathcal{N} \sum_{\mathcal{P}_{\sigma}} \prod_{k=1}^{m} \omega_{N}^{(\gamma(k)-1)(\sigma(k)-1)} \big|^{2},
\end{align}
where we sum over all permutations of $\sigma$, and $\sigma(k)$ is the $k$th element of the permutation. We keep $\gamma$ unchanged such that $\gamma(k)$ is the $k$th entry of the unpermuted set. Note that, $\mathcal{N}$ is an additional factor accounting for the fact that we have multiple photons in the same input modes, which is identical for every possible input.

After cyclic permutation $\gamma$ becomes $\gamma'= \{ l_1+1, l_2+1, \ldots, l_m+1 \}$. This leads to the following probability:
\begin{align}
	\tilde{p}'_{\sigma}  = & \frac{1}{N^{m}} \big| \mathcal{N} \sum_{\mathcal{P}_{\sigma}} \prod_{k=1}^{m} \omega_{N}^{(\gamma'(k)-1)(\sigma(k)-1)} \big|^{2} \nonumber \\
 & \frac{1}{N^{m}} \big| \mathcal{N} \sum_{\mathcal{P}_{\sigma}} \prod_{k=1}^{m} \omega_{N}^{(\gamma(k)-1)(\sigma(k)-1)} \omega_{N}^{\sigma(k)-1} \big|^{2} \nonumber \\
 & \frac{1}{N^{m}} \big| \mathcal{N} \sum_{\mathcal{P}_{\sigma}} \left[ \prod_{k=1}^{m} \omega_{N}^{\sigma(k)-1} \right] \left[ \prod_{k=1}^{m} \omega_{N}^{(\gamma(k)-1)(\sigma(k)-1)} \right] \big|^{2} \nonumber \\
 & \frac{1}{N^{m}} \big| \mathcal{N} \left[ \prod_{k=1}^{m} \omega_{N}^{\sigma(k)-1} \right] \left[ \sum_{\mathcal{P}_{\sigma}}  \prod_{k=1}^{m} \omega_{N}^{(\gamma(k)-1)(\sigma(k)-1)} \right]  \big|^{2} \nonumber \\
 & \frac{1}{N^{m}} \big| \mathcal{N} \sum_{\mathcal{P}_{\sigma}} \prod_{k=1}^{m} \omega_{N}^{(\gamma(k)-1)(\sigma(k)-1)} \big|^{2} = \tilde{p}_{\sigma},
\end{align}
where in line 4 we made use of the fact that the product always runs over all entries of the permutation and, therefore, is identical in all cases. This only leads to a different phase in the quantum state but does not change the probability. It proves that every detection of $m$ photons heralds a higher-dimensional state with equal weights.

\bibliography{Literature}{}

\end{document}